\documentstyle[prl,aps,epsf]{revtex} \textheight24cm

\begin{document}
\draft
\twocolumn\wideabs{
\title{ Resonant acousto-optics of microcavity polaritons }

\author{A.L. Ivanov}
\address{ Cardiff University, Department of Physics and Astronomy, 
Queen's Buildings, 5 The Parade, Cardiff CF24 3YB, 
Wales UK}

\author{P.B. Littlewood}
\address{University of Cambridge, Department of Physics, 
TCM Group, Cavendish Laboratory, Cambridge CB3 0HE, UK}

\date{December 6, 2002}
\maketitle
\begin{abstract}

We propose and analyze theoretically a resonant acousto-optic Stark effect 
for microcavity (MC) polaritons parametrically driven by a surface acoustic 
wave. For GaAs-based microcavities our scheme ``acoustic pumping - optical 
probing'' deals with surface acoustic waves of frequency $\nu_{\rm SAW} 
\simeq 0.5 - 3$\,GHz and intensity $I_{\rm SAW} \simeq 0.1 - 10$\,mW/mm. 
The acoustically-induced stop gaps in the MC polariton spectrum drastically 
change the optical response of MC polaritons. Because an acoustically 
pumped intrinsic semiconductor microcavity remains in its ground electronic 
state, no many-body effects screen and weaken the resonant acousto-optic 
Stark effect. In the meantime, this allows us to work out an 
exactly-solvable model for resonant acousto-optics of MC polaritons which 
deals with giant acousto-optical nonlinearities. Finally, we discuss 
possible applications of the proposed resonant acoustic Stark effect for 
optical modulation and switching and describe an acousto-optic device based 
on a (GaAs) microcavity driven by a surface acoustic wave.

\end{abstract}
}

\narrowtext

\section{Introduction}

The science of acousto-optics is a well-established field that relies on 
the modulation of the dielectric constant of a material by a pattern of 
sound waves to produce diffraction of incident light 
\cite{Sapriel79,Wilson83,Korpel97}. In the conventional acousto-optics, 
the principle effect on the dielectric constant arises just from the 
modulation of the density; correspondingly the modulation is weak and 
deals with small nonresonant acousto-optical nonlinearities. Most of the 
light on the sample will pass through undiffracted. 

The resonant acousto-optics \cite{Ivanov01a,Ivanov01b} we develop here for 
microcavity polaritons associated with quantum well (QW) excitons 
deals with giant acousto-optical nonlinearities mediated by the QW 
excitonic resonance. In this case the momentum-energy matching condition, 
necessary for appearance of the acoustically-induced optical stop gaps, 
occurs for optical frequencies nearly resonant with QW excitons. As a 
result, the acousto-optical susceptibility is enhanced by several orders 
of magnitude in comparison to conventional applications, both due to 
small detuning from the excitonic resonance and the relatively small 
damping constants relevant to QW excitons and MC photons (for GaAs-based 
MCs a typical polaritonic linewidth is less than 0.5\,meV). 

Linear and nonlinear optics of MC polaritons are now a well-established 
field (see for example  
\cite{Weisbuch92,Savasta96,Dang98,Savvidis00,Boeuf00,Ciuti00,Borri00, 
Baumberg02,Pawlis02}). 
There are several points in favour of the resonant acousto-optics of MC 
polaritons: (i) a well-controlled in-plane wave interaction in 
microcavities, (ii) a compatibility of semiconductor microcavities with 
surface acoustic wave (SAW) technique, and (iii) a possibility to realize 
one-dimensional geometry for resonant, SAW-mediated interaction of two 
counter-propagating MC polaritons.  

In this article, we discuss how SAW and microcavity technologies can be 
combined to make devices where the SAW can produce a tunable and narrow 
optical band gap, while being driven at modest powers. An attractive feature 
of this technique is that the process does not involve real excitation of 
excitons or electron-hole pairs. The strong resonant coupling produces an 
interaction length of typically 100\,$\mu$m or less, making the process 
suitable for device applications. Our analysis is provided here for 
appropriate parameters for GaAs at low temperatures, since this is the 
material for which both the technologies of SAW and microcavity are well 
developed.  However, the phenomenon occurs in any material with a sharp 
resonance in the optical spectrum and a strong piezo-electric 
or deformation potential coupling; which includes organic semiconductors, 
other semiconductors such as CdTe or ZnSe, and rare-earth-doped glass. In 
many of these materials, the effects are much larger than in GaAs, and 
will be substantial even at room temperature.

The discovery of the optical Stark effect mediated by an excitonic 
resonance in semiconductors and semiconductor nanostructures 
\cite{Froehlich85,VonLehmen86,Mysyrowicz86} initiated numerous attempts 
to apply the effect for ultra-fast light-light processing. Later on, 
however, it was realized that the excitonic optical Stark effect requires 
high-intensity laser fields, $I_{\rm opt} \simeq 0.1 - 10\,\mbox{GW/cm}^2$, 
so that for a pump electromagnetic pulse nearly resonant with excitons 
it is practically impossible to avoid generation of real, incoherent 
electrons and holes, keeping a semiconductor under purely virtual optical 
excitation. As a result, the incoherent scattering rate of excitons 
strongly increases with increasing $I_{\rm opt}$ and considerably relaxes 
the optically-induced spectral changes. Furthermore, a high-intensity laser 
pulse, which gives rise to the optical Stark effect, weakens the 
exciton-photon interaction and relaxes the excitonic states, due to the 
Coulomb interaction between photogenerated virtual (and real) excitons 
and due to Pauli blocking of the relevant single-electron states. The 
above arguments explain why the exciton-mediated optical Stark effect, 
which refers to the scheme ``optical pumping - optical probing'', 
cannot practically be used for optoelectronic applications. 

By the same arguments, particular realizations of the exciton-mediated 
optical Stark effect, due to excitonic molecules 
\cite{Combescot88,Ivanov89,Hulin90,Shimano94} or due to exciton-phonon 
interaction \cite{Ivanov86,Greene88,Kiseleva99}, remain mainly a question 
of interesting fundamental science with little practical application as 
yet. For example, the acoustic phonoriton spectrum, which gives rise to 
the phonon-mediated Stark effect, has recently been measured and analyzed 
for orthoexcitons in Cu$_2$O driven by an optical field of intensity 
$I_{\rm opt} \simeq 0.1 - 1$~GW/cm$^2$ \cite{Hanke99,Hanke00}. In spite of 
relatively strong coupling of excitons with acoustic phonons in GaAs, 
one cannot realize the phonon-mediated optical Stark effect in this 
semiconductor: the large Bohr radius of excitons does not allow the use of 
high $I_{\rm opt}$ without the modification and eventual removal of the 
excitonic states. 

The resonant acousto-optics, which in this paper we develop for MC 
polaritons, deals with an alternative scheme of the dynamic Stark effect: 
acoustically driven MC polaritons, associated with QW excitons, probed by 
weak light (acoustically-induced Stark effect). Because an acoustically 
pumped intrinsic semiconductor microcavity remains in its ground electronic 
state, no many-body effects screen the resonant acousto-optic Stark effect 
and weaken the QW exciton - MC photon interaction. Being applied to 
GaAs-based MCs, the proposed scheme ``acoustic pumping - optical probing'' 
effectively exploits both the relatively strong exciton-phonon piezoelectric 
interaction and the considerable exciton-photon coupling relevant to bulk 
GaAs \cite{Ulbrich77}. As we show below, the above processing scheme 
requires moderate-intensity acoustic and low-intensity electromagnetic waves 
but can still operate in a picosecond time domain (the latter needs a 
particular design of a GaAs microcavity driven by a surface acoustic wave, 
see section 4). Thus the resonant acousto-optic Stark effect for MC 
polaritons takes all the advantages of the excitonic optical Stark effect, 
but needs much smaller operating intensities. 

In the last decade SAWs were widely applied mainly to study electron and 
hole transport in GaAs-based QW structures. Surface acoustic waves with 
frequency $\nu_{\rm SAW} = 2-9$\,GHz have been successfully used to 
investigate the fractional quantum Hall effect for a quasi-two-dimensional 
electron system in low-temperature GaAs/AlGaAs heterostructures 
\cite{Willett93a,Willett93b}. A single-electron acoustoelectric current 
induced by SAWs with $\nu_{\rm SAW} = 2-3$\,GHz in a quasi-one-dimensional 
(quasi-1D) GaAs channel has been observed and analyzed 
\cite{Shilton96,Cunningham00}. Acoustically-induced ionization of 
photogenerated QW excitons and subsequent long-distance acoustoelectric 
transport of free electrons and holes have been detected in InGaAs/GaAs 
QWs driven by a SAW with $\nu_{\rm SAW} = 0.1-0.9$\,GHz 
\cite{Rocke97,Rocke98,Govorov00}. The surface acoustical waves with 
experimentally accessible parameters, frequency $\nu_{\rm SAW} \lesssim 
2-3\,$GHz and moderate intensity $I_{\rm SAW} \lesssim 10$\,mW/mm, are also 
very suitable for effective realization of the resonant acousto-optics of 
MC polaritons. Note that in this case the SAW intensities are much less 
than those used to ionized the QW excitons \cite{Rocke97,Rocke98}.

\section{Microcavity polaritons parametrically driven by the 
coherent acoustic field}

Our model deals with a QW microcavity parametrically driven by an in-plane 
propagating coherent acoustic wave of frequency $\Omega^{\rm ac}_{\bf k}$, 
wavevector ${\bf k}$ and intensity $I_{\rm ac}$. In this case the 
Hamiltonian of QW excitons and in-plane MC photons is given by 
\begin{eqnarray}
H &=& \sum_{{\bf p}_{\|}} \Bigg[  \hbar \omega^{\rm X}_{{\bf p}_{\|}}
B_{{\bf p}_{\|}}^{\dag} B_{{\bf p}_{\|}} + \hbar 
\omega^{\rm MC}_{{\bf p}_{\|}} \alpha_{{\bf p}_{\|}}^{\dag} 
\alpha_{{\bf p}_{\|}} 
\nonumber \\ 
&+& i \hbar { \Omega_{\rm X}^{\rm MC} \over 2 } 
\Big( \alpha_{{\bf p}_{\|}}^{\dag} B_{{\bf p}_{\|}} - 
B_{{\bf p}_{\|}}^{\dag} \alpha_{{\bf p}_{\|}}\Big) 
\nonumber \\ 
&+& \ i m^{\rm X}_{\bf k} \left( B_{{\bf p}_{\|}}^{\dag} 
B_{{\bf p}_{\|}-{\bf k}} e^{-i \Omega^{\rm ac}_{\bf k} t} 
-  B_{{\bf p}_{\|}-{\bf k}}^{\dag} B_{{\bf p}_{\|}} 
e^{i \Omega^{\rm ac}_{\bf k} t} \right) \Bigg] \ , 
\eqnum{1} 
\label{ham}
\end{eqnarray}
where ${\bf p}_{\|}$ is the in-plane wavevector, $B_{{\bf p}_{\|}}$ are 
$\alpha_{{\bf p}_{\|}}$ are the QW exciton and MC photon operators, 
respectively, $\hbar \omega^{\rm X}_{{\bf p}_{\|}} = \hbar \omega_t + 
(\hbar^2 p_{\|}^2)/(2M_x)$ and $\hbar \omega^{\rm MC}_{{\bf p}_{\|}}  
= \hbar \sqrt{ \omega_0^2 + c^2 p^2_{\|}/\varepsilon_b}$ are the 
corresponding energies, $M_x$ is the QW exciton translational mass, 
$\omega_0$ is the cavity (photon) eigenfrequency, and $\varepsilon_b$ is 
the background dielectric constant. The oscillator strength of the 
resonant coupling between QW excitons and MC photons is characterized by 
the polariton Rabi frequency, $\Omega_{\rm X}^{\rm MC}$. The parametric 
interaction of MC polaritons with the acoustic pump wave occurs through 
their excitonic component and is given by $m_{\bf k}^{\rm X} = 
m_{\rm x-ac} \sqrt{N_0^{\rm ph}}$. Here $m_{\rm x-ac}$ is the matrix 
element of exciton-phonon interaction and $N_0^{\rm ph} \propto 
I_{\rm ac}$ is the concentration of the coherent acoustic phonons. For 
definiteness we assume a (GaAs) $\lambda$-microcavity so that $\omega_0 = 
(2 \pi c)/(L_z \sqrt{\varepsilon_b})$, where $L_z$ is the MC thickness. 

As we discuss in detail below, the SAW properties and the SAW generation 
technique are very suitable for the effective realization of the resonant 
acousto-optic effect in optical microcavities. The interaction of QW 
excitons with SAWs is piezoelectric so that the matrix element 
$m_{\bf k}^{\rm X}$ is given by 
\begin{equation}
m_{\bf k}^{\rm X} = \left( { 2 N_0^{\rm ph} \hbar k \over \rho 
v_s} \right)^{1/2} { \pi e \over \varepsilon_b } \, \mbox{e}_{14} 
k a_{\rm B}^2 \left( {m_{\rm e} - m_{\rm h} \over M_x} \right) \, , 
\eqnum{2} 
\label{matr}
\end{equation}
where $\rho$ is the mass density, $v_s \equiv v_{\rm SAW}$ is the sound 
velocity, ${\rm e}_{14}$ is the relevant component of an electromechanical 
tensor of the electron - phonon piezoelectric interaction, $a_{\rm B}$ is 
the QW exciton Bohr radius, and $m_{\rm e(h)}$ is the QW electron (hole) 
mass. Equation (\ref{matr}) is valid for $(a_{\rm B} k)/2 \ll 1$. For 
GaAs-based structures this means that the wavevector $k$ of the SAW should 
be less than $5 \times 10^5\,\mbox{cm}^{-1}$. For GaAs one has ${\rm e}_{14} 
= 0.48 \times 10^5\,\mbox{g}^{1/2}\mbox{s}^{-1}\mbox{cm}^{-1/2}$ \cite{Yu96} 
and $v_{\rm SAW} \simeq 2.87 \times 10^5$\,cm/s \cite{Rocke97}. The latter 
value clearly shows that the SAW velocity is less than the sound velocities 
of bulk, transverse and longitudinal, acoustic phonons. The concentration 
of the coherent phonons $N_0^{\rm ph}$, associated with the SAW, is given 
by  $N_0^{\rm ph} = I_{\rm ac}/(v_s \hbar \Omega_{\bf k}^{\rm ac})$. Here 
the phonon eigenfrequency $\Omega_{\bf k}^{\rm ac}= v_s k$ is resonant to 
the angular SAW frequency $\Omega_{\rm SAW}$, i.e. $\Omega_{\bf k}^{\rm ac} 
= \Omega_{\rm SAW}$. Usually the SAW frequency is given in terms of 
$\nu_{\rm ac} \equiv \nu_{\rm SAW} = \Omega_{\rm SAW}/(2 \pi)$ so that 
the phonon wavelength is $\lambda_{\rm ac} \equiv \lambda_{\rm SAW} = 
v_{\rm SAW}/\nu_{\rm SAW}$. 

The explicit time dependence of the Hamiltonian (\ref{ham}), due to 
the cw acoustic pump wave, can be removed by the canonical transformation: 
\begin{eqnarray}
&&S = \exp \Big[ it \sum_{{\bf p}_{\|}} ({\bf v_s} \cdot {\bf p}_{\|}) 
\left( B_{{\bf p}_{\|}}^{\dag} B_{{\bf p}_{\|}}  + 
\alpha_{{\bf p}_{\|}}^{\dag} \alpha_{{\bf p}_{\|}} \right) \Big] \, ,
\nonumber \\
&&B_{{\bf p}_{\|}} \rightarrow S B_{{\bf p}_{\|}} S^{\dag} = 
B_{{\bf p}_{\|}} e^{-i({\bf v_s} \cdot {\bf p}_{\|})t}  \, , 
\nonumber \\
&&\alpha_{{\bf p}_{\|}} \rightarrow S \alpha_{{\bf p}_{\|}} S^{\dag} = 
\alpha_{{\bf p}_{\|}} e^{-i({\bf v_s} \cdot {\bf p}_{\|})t}  \, ,
\eqnum{3} 
\label{canon}
\end{eqnarray}
where ${\bf v_s} = v_s ({\bf k}/k)$ and, therefore, $({\bf v_s} \cdot 
{\bf p}_{\|}) = \Omega_{\bf k}^{\rm ac} [ ({\bf k} \cdot {\bf p}_{\|})/k^2 ]$. 
The canonical transformation (\ref{canon}) means the use of the coordinate 
system which moves with the acoustic wave. In this case the quadratic 
Hamiltonian ${\tilde H} = S H S^{\dag} - i S (\partial S^{\dag} / 
\partial t)$ is time-independent at least insofar as the rotating wave 
approximation made in equation (1) is respected. The Hamiltonian 
${\tilde H}$ is then exactly-solvable. Its diagonalization yields the 
dispersion equation for MC polaritons parametrically driven by the coherent 
(surface) acoustic wave:
\begin{eqnarray}
\omega^2 &-& ({\tilde \omega}^{\rm X}_{{\bf p}_{\|}})^2 - 
{ ( \omega \Omega_{\rm X}^{\rm MC} )^2 \over \omega^2 - 
({\tilde \omega}^{\rm MC}_{{\bf p}_{\|}})^2 } 
\nonumber \\ 
&-& { 4 \omega_t^2 |m_{\bf k}^{\rm X}|^2 \over 
\omega^2 - ({\tilde \omega}^{\rm X}_{{\bf p}_{\|}+{\bf k}})^2 
- { ( \omega \Omega_{\rm X}^{\rm MC} )^2 \over \omega^2 - 
({\tilde \omega}^{\rm MC}_{{\bf p}_{\|}+{\bf k}})^2 } 
- M_{{\bf p}_{\|}+2{\bf k}} } 
\nonumber \\ 
&-& { 4 \omega_t^2 |m_{\bf k}^{\rm X}|^2 \over 
\omega^2 - ({\tilde \omega}^{\rm X}_{{\bf p}_{\|}-{\bf k}})^2 
- { ( \omega \Omega_{\rm X}^{\rm MC} )^2 \over \omega^2 - 
({\tilde \omega}^{\rm MC}_{{\bf p}_{\|}-{\bf k}})^2 } 
- M_{{\bf p}_{\|}-2{\bf k}} } = 0 \, , 
\eqnum{4} 
\label{disp}
\end{eqnarray}
where $M_{{\bf p}_{\|} \pm 2{\bf k}}$ is given through $M_{{\bf p}_{\|} 
\pm 3{\bf k}}$, $M_{{\bf p}_{\|} \pm 3{\bf k}}$ through $M_{{\bf p}_{\|} 
\pm 4{\bf k}}$, etc. by the recurrent formula valid for a positive 
integer $n \geq 1$: 
\begin{equation}
M_{{\bf p}_{\|} \pm n{\bf k}} = { 4 \omega_t^2 |m_{\bf k}^{\rm X}|^2 \over 
\omega^2 - ({\tilde \omega}^{\rm X}_{{\bf p}_{\|} \pm n{\bf k}})^2 
- { ( \omega \Omega_{\rm X}^{\rm MC} )^2 \over \omega^2 - 
({\tilde \omega}^{\rm MC}_{{\bf p}_{\|} \pm n{\bf k}})^2 } 
- M_{{\bf p}_{\|} \pm (n+1){\bf k}} } 
\eqnum{5} 
\label{recur}
\end{equation}
and $\hbar {\tilde \omega}^{\rm X,MC}_{{\bf p}_{\|} \pm n{\bf k}} = \hbar 
\omega^{\rm X,MC}_{{\bf p}_{\|} \pm n{\bf k}} \mp n \hbar 
\Omega_{\bf k}^{\rm ac} = \hbar \omega^{\rm X,MC}_{{\bf p}_{\|} \pm 
n{\bf k}} \mp n \hbar ({\bf v_s} \cdot {\bf k})$ are the QW exciton and MC 
photon quasi-energies, respectively. 

In the derivation of the dispersion equation (\ref{disp}) we have treated 
the interaction between QW excitons and MC photons in its general, 
nonresonant approximation. If $|m_{\bf k}^{\rm X}|^2 \propto I_{\rm ac}$ 
is equal to zero, i.e. in the absence of the acoustic pump wave, 
equation (\ref{disp}) reduces to the MC polariton dispersion relationship:  
\begin{equation}
{ c^2 p_{\|}^2 \over \varepsilon_b } + \omega^2_0 = \omega^2 
+ { \omega^2 (\Omega^{\rm MC}_{\rm X})^2 \over \omega^2_t + 
\hbar \omega_t p_{\|}^2/M_x - \omega^2 \ } \ .   
\eqnum{6}
\label{MC}
\end{equation}
In this case the exciton and photon components of MC polaritons, 
$\varphi^{\rm MC}$ and $\psi^{\rm MC}$, are given by 
\begin{eqnarray}
&& \varphi^{\rm MC}(p_{\|}) =  { (\Omega^{\rm MC}_{\rm X})^2 \over  
(\Omega^{\rm MC}_{\rm X})^2 + 4 [\omega_t  + \hbar  p_{\|}^2/(2M_x) -  
\omega_{\rm pol}^{\rm MC}(p_{\|}) ]^2 } \ , 
\nonumber \\
&& 
\psi^{\rm MC}(p_{\|}) = 1 - \varphi^{\rm MC}(p_{\|}) \ , 
\eqnum{7}
\label{comp}
\end{eqnarray}
respectively, where $\omega_{\rm pol}^{\rm MC} = 
\omega^{\rm MC}_{\rm LPB(UPB)}(p_{\|})$ is the MC lower (upper) 
dispersion branch determined by equation (\ref{MC}). 

The dispersion equation (\ref{disp}) describes the resonant acousto-optic 
(Stark) effect for MC polaritons. The equation is suitable for numerical 
calculations within a truncated scheme, i.e. by putting $M_{{\bf p}_{\|} \pm 
n{\bf k}}$ equal to zero at some integer $n$=$n^{\rm max} > 1$. The truncated 
scheme quickly converges with increasing $n^{\rm max}$. According to equation 
(\ref{disp}), with increasing $I_{\rm ac}$ the optical band gaps (optical 
stop gaps) open up and develop in the MC polariton spectrum. The gaps 
arise for the MC polariton states resonantly coupled by one-, two-, etc. 
phonon-assisted transitions induced by the acoustic pump wave. The 
quasi-energy spectrum (\ref{disp}) can be interpreted in terms of the 
Brillouin energy bands: the initial MC polariton dispersion is modulated by 
the periodic potential associated with the coherent acoustic wave. The 
latter potential is periodic not only in space, with the ``lattice'' 
constant $\lambda_{\rm ac}$, but also in time, with the period 
$1/\nu_{\rm ac}$. This gives rise to the quasi-energies $\hbar 
{\tilde \omega}^{\rm X,MC}_{{\bf p}_{\|} \pm n{\bf k}}$ in equation 
(\ref{disp}) (the concept of quasi-energies has been introduced and 
developed in \cite{Zeldovich66}). Thus not only two in-plane momenta, $\hbar 
{\bf p}_{\|}$ and $\hbar {\bf p}_{\|} \pm n \hbar {\bf k}$, are equivalent 
for acoustically-driven MC polaritons, but in a similar way two energies, 
$\hbar \omega^{\rm X,MC}_{{\bf p}_{\|}}$ and $\hbar 
\omega^{\rm X,MC}_{{\bf p}_{\|} \pm n{\bf k}} \mp n \hbar 
\Omega_{\bf k}^{\rm ac}$, are also indistinguishable. 

The quasi-energy spectrum of bulk polaritons driven by a bulk acousic pump 
wave has been discussed in detail in \cite{Ivanov01a}. Here we concentrate 
on the first (main) optical stop gap associated with the one-phonon 
resonant transition between the MC polariton states. In figure\,1(a) we plot 
the initial MC polariton dispersion calculated by equation (\ref{MC}) for 
zero detuning at $p_{\|}=0$ between the cavity mode and the QW exciton 
energy, i.e. for $\omega_0 = \omega_t$. The parameters relevant to 
GaAs-based microcavities have been used in the calculations: the Rabi 
energy $\hbar \Omega^{\rm MC}_{\rm X} = 3.7$\,meV, the dielectric constant 
$\varepsilon_b = 12.3$ and the QW exciton energy $E_{\rm X} = \hbar \omega_t 
= 1.522$\,eV. The applied SAW is characterized by the wavevector $k = 5 
\times 10^4\,\mbox{cm}^{-1}$ and the frequency $\nu_{\rm ac} = 2.28$\,GHz 
(see figure\,1(a), where only the one-phonon-assisted transition within the 
lower MC polariton dispersion branch is indicated by the arrow 1LPLP). 
Figure\,1(b) illustrates how the first optical band gap develops with 
increasing acoustic intensity $I_{\rm ac} \equiv I_{\rm SAW}$. The plot has 
been calculated by equation (\ref{disp}) using the truncated scheme with 
$n^{\rm max} = 4$. 

The first (main) optical band gap is due to the MC polariton states 
$\{ {\bf p}_{\|}, \omega^{\rm MC}_{\rm LPB}({\bf p}_{\|}) \}$ and 
$\{ {\bf p}_{\|}-{\bf k}, \omega^{\rm MC}_{\rm LPB}({\bf p}_{\|}-{\bf k}) 
\}$ resonantly coupled by the one-phonon-assisted transition induced by 
the SAW $\{ {\bf k},\Omega_{\bf k}^{\rm ac}\equiv \Omega_{\rm SAW} \}$ 
(see figures\,1(a) and 1(b)). Because the energy of a SAW phonon is 
rather small, $\hbar \Omega_{\rm SAW} \simeq 9.4\,\mu$eV, for the 
one-dimensional geometry, ${\bf p}_{\|} \, \| \, {\bf k}$, one has that 
the momentum-energy matching condition is nearly satisfied for the MC 
states $\{ {\bf p}_{\|}, \omega^{\rm MC}_{\rm LPB}({\bf p}_{\|}) \} 
= \{ {\bf k}/2, \omega^{\rm MC}_{\rm LPB}(k/2) \}$ and $\{ {\bf p}_{\|} - 
{\bf k},\omega^{\rm MC}_{\rm LPB}({\bf p}_{\|}-{\bf k}) \} = 
\{ -{\bf k}/2, \omega^{\rm MC}_{\rm LPB}(k/2) \}$. Thus, using 
equation (\ref{disp}), we conclude that the first optical band gap 
$\Delta^{\rm MC}_{\rm ac} \equiv \Delta^{{\rm MC}(n=1)}_{\rm ac} \propto 
\sqrt{I_{\rm ac}}$ is given approximately by 
\begin{eqnarray}
\Delta^{\rm MC}_{\rm ac} &=& \Delta_{\rm ac}^{\rm X}(I_{\rm ac}) 
[\varphi^{\rm MC}({\bf k}/2) \varphi^{\rm MC}(-{\bf k}/2)]^{1/2} 
\nonumber \\ 
&\equiv&  \Delta_{\rm ac}^{\rm X}(I_{\rm ac}) \varphi^{\rm MC}(k/2) \, , 
\eqnum{8}
\label{gap}
\end{eqnarray} 
where $\Delta_{\rm ac}^{\rm X} = 2 |m_{\bf k}^{\rm X}| \propto 
\sqrt{I_{\rm ac}}$ is the stop band for optically-undressed QW excitons, 
and $\varphi^{\rm MC} = \varphi^{\rm MC}_{\rm LPB}$ is given by equation 
(\ref{comp}). For the piezoelectric interaction, which is determined 
by equation (\ref{matr}), one has that $\Delta_{\rm ac}^{\rm X}$ is 
proportional to the phonon wavevector $k$. According to equation (\ref{gap}), 
the acoustically created gap in the MC polariton spectrum can also be 
interpreted in terms of coherent phonon-mediated coupling of virtual, 
optically-dressed QW excitons with the in-plane wavevectors ${\bf k}/2$ 
and $-{\bf k}/2$. In the time domain, the optical band gap can be visualized 
as acoustically induced coherent oscillations of MC polaritons, forth 
and back between the states ${\bf k}/2$ and $-{\bf k}/2$. The frequency 
of the oscillations is given by $\Delta^{\rm MC}_{\rm ac}$. 

An acoustic intensity threshold for the resonant acousto-optic Stark 
effect is approximately given by the condition 
$\Delta^{\rm MC}_{\rm ac}(I_{\rm ac}) \geq \mbox{Max} \{ \Gamma_{\rm R}/2, 
\gamma_{\rm X}/2 \}$, where $\Gamma_{\rm R}$ is the inverse radiative 
lifetime of MC photons and $\gamma_{\rm X}$ is the rate of incoherent 
scattering of QW excitons. The rates have to be included in the dispersion 
equation (\ref{disp}) by the substitution $\omega^2_t \rightarrow \omega^2_t - 
i \omega \gamma_{\rm X}$ and  $\omega^2_0 \rightarrow \omega^2_0 - i 
\omega \Gamma_{\rm R}$. Note that both damping constants, $\Gamma_{\rm R}$ 
and $\gamma_{\rm X}$, refer to the decay rates of the occupation numbers of 
the relevant in-plane MC photon and QW exciton modes. The corresponding 
$T_1$-times are given by $T_1^{\rm R} = \hbar/\Gamma_{\rm R}$ and 
$T_1^{\rm X} = \hbar/\gamma_{\rm X}$, respectively, and we assume that 
the $T_2 = 2T_1$ limit holds for MC photons and QW excitons. A more precise 
condition for the threshold $I_{\rm ac}$ is 
\begin{equation}
\Delta^{\rm MC}_{\rm ac}(I_{\rm ac},{\bf k}) \geq {1 \over 2} \mbox{Max} \,  
\left\{
\begin{array}{ll}
\psi^{\rm MC}(k/2) \Gamma_{\rm R} \, , \\ 
\varphi^{\rm MC}(k/2) \gamma_{\rm X} \, , \\ 
\end{array}
\right.
\eqnum{9}
\label{thresh}
\end{equation}
where $\psi^{\rm MC}$ and $\varphi^{\rm MC}$ are given by equation 
(\ref{comp}). Using equation (\ref{gap}), the second inequality can also 
be re-written as $\Delta^{\rm X}_{\rm ac} \geq \gamma_{\rm X}/2$. For the 
up-to-date GaAs-based microcavites at helium temperatures, $\Gamma_{\rm R}$ 
is much larger than $\gamma_{\rm X}$. For example, for high-quality GaAs 
MCs the radiative lifetime of MC photons $T_1^{\rm R}$ is about 
$2-3$\,ps \cite{Borri00} so that $\hbar \Gamma_{\rm R} \simeq 0.3$\,meV. 
In the meantime, at $T=4.2$\,K the scattering rate of QW excitons yields 
$\hbar \gamma_{\rm X} \simeq 30-60\,\mu$eV \cite{Ivanov99,Nagerl01}. The 
latter values refer to the low density limit of QW excitons. Note that we 
do have this limit for the resonant acousto-optic Stark effect: our scheme 
does not require high optical excitations, while the acoustic pump wave 
cannot generate the QW excitons itself in intrinsic, undoped nano-structures. 
In contrast, the conventional optical Stark effect, mediated by excitons, 
needs a high-intensity electromagnetic wave which nearly resonates with the 
interband electronic transitions and, therefore, inevitably populates the 
excitonic and free electron (hole) states. As a result, $\gamma_{\rm X}$ 
considerably increases in an uncontrolled way and relaxes the optical 
Stark effect.

From the dispersion equation (\ref{disp}) we find the acoustically-mediated 
excitonic susceptibility for MC polaritons: 
\begin{eqnarray}
\chi^{\rm MC}_{\rm x-ac}(I_{\rm ac}) &=& - { \varepsilon_b 
\over 4 \pi } \, { (\Omega^{\rm MC}_{\rm X})^2
\over \omega^2 - ({\tilde \omega}^{\rm X}_{{\bf p}_{\|}})^2  }
\nonumber \\ 
&\times& { M_{{\bf p}_{\|}+{\bf k}} + M_{{\bf p}_{\|}-{\bf k}}
\over \omega^2 - ({\tilde \omega}^{\rm X}_{{\bf p}_{\|}})^2  - 
M_{{\bf p}_{\|}+{\bf k}} - M_{{\bf p}_{\|}-{\bf k}}   } \, , 
\eqnum{10}
\label{susc}
\end{eqnarray}
where $M_{{\bf p}_{\|} \pm {\bf k}}$ is given by equation (\ref{recur}) 
with $n=1$. The susceptibility $\chi^{\rm MC}_{\rm x-ac}$, which is linear 
with respect to the light field, can also be interpreted as a resonant 
acousto-optical nonlinear susceptibility 
$\chi^{\rm MC}_{\rm x-ac}(I_{\rm ac}) = \sum_{n=1}^{\infty} 
\chi_{\rm x-ac}^{(2n+1)} I_{\rm ac}^n$. Note that $\chi^{\rm MC}_{\rm x-ac}$ 
is strongly enhanced due to small detuning $\omega - \omega_t$ from the QW 
excitonic resonance and due to the small values of $\Gamma_{\rm R}$ and 
$\gamma_{\rm X}$. According to equation (\ref{susc}), the 
acoustically-induced change of the dielectric constant $\varepsilon_b$, 
associated with the third-order $\chi^{(3)}_{\rm x-ac}$, has a maximum value 
given by 
\begin{eqnarray}
{ | \delta \varepsilon_b^{(3)}(\omega,I_{\rm ac}) |
\over \varepsilon_b } \bigg|_{\rm max}  &=& 
{ (\Omega_{\rm X}^{\rm MC})^2 \varphi^{\rm MC} \over 2 \omega_t 
( \varphi^{\rm MC} \gamma_{\rm X} + \psi^{\rm MC} \Gamma_{\rm R} ) } 
\nonumber \\
&\times& 
{ (\Delta_{\rm ac}^{\rm X})^2 \over ( \omega - \omega_t )^2 } \propto 
I_{\rm ac} \, , 
\eqnum{11}
\label{huge}
\end{eqnarray}
where the light field of frequency $\omega$ resonates with the centre of the 
acoustically-induced band gap, and we assume that $|\omega - \omega_t| \gg 
\gamma_{\rm X}/2$. For example, for the parameters relevant to figure\,1(b) 
one has $\omega - \omega_t \simeq - 0.49$\,meV, 
$\varphi^{\rm MC}_{\rm LPB}(\omega) \simeq 0.93$ and 
$\psi^{\rm MC}_{\rm LPB}(\omega) \simeq 0.07$ so that equation (\ref{huge}) 
yields $| \delta \varepsilon_b^{(3)}(I_{\rm SAW} = 4.5\,\mbox{mW/mm}) 
|_{\rm max} \simeq 0.54$. The latter value clearly shows that the resonant 
acousto-optics of MC polaritons deals with the giant acousto-optical 
nonlinearities.

\section{Discussion}

If the incoherent scattering processes of QW excitons and the radiative 
decay rates of MC polaritons are neglected, the acoustically-induced stop 
gaps in the MC polariton spectrum correspond to a real and negative 
effective dielectric constant $\varepsilon^{\rm eff} = 
\varepsilon^{\rm eff}(\omega,I_{\rm ac})$ and yield total reflection for 
the weak light field which activates and probes the acoustically-driven MC 
polaritons. In this case sharp spikes, which are associated with the 
acoustically-induced optical stop gaps, arise and develop with increasing 
$I_{\rm ac}$ in the reflection spectrum of acoustically-dressed MC polaritons 
(the reflection spectrum of acoustically-driven excitons in bulk GaAs is 
shown in \cite{Ivanov01a}). The main acoustically-induced spike corresponds 
to the one-phonon resonant coupling between two MC polariton states (see 
figures 1(a) and 1(b)). The contrast of the spikes decreases with the 
increasing scattering rates $\gamma_{\rm X}$ and $\Gamma_{\rm R}$, and finally 
the acoustically-induced changes of the reflection spectrum disappears 
when $\Delta^{\rm MC}_{\rm ac} = \Delta^{\rm MC}_{\rm ac}(I_{\rm ac})$ 
becomes much less than Max$\{ \Gamma_{\rm R}/2,\gamma_{\rm X}/2 \}$. Note 
that the spectral position of the acoustically-induced band gaps can 
easily be tuned by sub-GHz changes in the frequency 
$\Omega^{\rm ac}_{\bf k}$ of the acoustic pump wave. 

The standard acousto-optic effect can be interpreted in terms of optical 
diffraction of the incident light field by the acoustically-induced grating 
of the period $\lambda_{\rm ac}$. In this case a coherent acoustic pump 
wave can be treated as nonpropagating, because $c \gg v_s$. The 
acoustically-induced diffraction grating is due to the photoelastic effect, 
which gives rise to a small, nonresonant change of the dielectric constant, 
$|\delta \varepsilon_b| \ll \varepsilon_b$, at acoustic wave peaks (pressure 
maxima so that usually $\delta \varepsilon_b$ is positive) and troughs 
(pressure minima so that usually $\delta \varepsilon_b$ is negative). In 
sharp contrast with the above picture, the resonant acousto-optic effect 
we describe deals with a quantum diffraction of optically-active MC 
excitons by an ultrasonic coherent wave. In this case the interaction of two 
``matter'' waves, the excitonic polarization and the acoustic field, is much 
more effective than nonresonant coupling of photons with acoustic phonons. 
In the meantime the photon component of optically-dressed MC excitons, i.e. 
of MC polaritons, can be large enough to ensure an efficient resonant 
conversion ``MC photon $\leftrightarrow$ MC exciton''. This explains 
qualitatively the origin of the giant acousto-optic nonlinearities
$4 \pi \chi_{\rm x-ac}^{\rm MC} \gg \delta \varepsilon_b$, which are mediated 
by the exciton states nearly resonant with the acoustic and optic fields 
simultaneously. 

For the frequency band of visible light the one-dimensional geometry of 
the acousto-optic interaction can hardly be realized within the nonresonant 
acousto-optic effect in bulk materials, because in order to couple two 
mutually-opposite states of the photon cone, ${\bf p}$ and ${\bf p} - 
{\bf k} \simeq - {\bf p}$, one needs an acoustic frequency $\nu_{\rm ac} 
\gtrsim 20\,$GHz. The up-to-date SAW technique deals with $\nu_{\rm ac} 
\lesssim 10\,$GHz. Thus usually the photon wavevector $|{\bf p}|$ is much 
larger than $|{\bf k}| = 2\pi/\lambda_{\rm ac}$, so that only a nearly normal 
interaction geometry, ${\bf p} \perp {\bf k}$, is used in the standard 
Raman-Nath (transmission type) and Bragg (reflection type) acousto-optic 
diffraction grating schemes applied to a bulk medium \cite{Wilson83,He89}. 
The 1D interaction geometry can easily be realized for the resonant 
acousto-optic effect in optical microcavities. Namely, due to a particular 
shape of the MC polariton dispersion branches with the closely separated 
absolute energy minima at ${\bf p}_{\|} = {\bf 0}$, a pump ultrasonic 
acoustic wave of any frequency $\nu_{\rm ac} \lesssim 2-3\,$GHz couples two 
counter-propagating MC polaritons with wavevectors ${\bf p}_{\|} \simeq \pm 
{\bf k}/2$ (see figure 1(a)). By the same reason the 1D geometry of the 
phonon-mediated interaction between counter-propagating photons 
can be realized by using the nonresonant acousto-optical nonlinearities 
in SAW-driven optical fibers. 

Due to the small values of the nonresonant acousto-optical nonlinearities, 
the interaction length between the photon and phonon fields, $l_{\gamma - 
{\rm ac}}$, needed for the usual acousto-optic effect, is large. For 
example, for the SAW-driven optical fibers, $l_{\gamma - {\rm ac}}$ is on a 
scale of tens of centimeters \cite{Campbell89,Ostling95}. In sharp contrast 
with this, the resonant acousto-optic Stark effect for MC polaritons 
requires the interaction length $l_{\rm MC-ac} \sim 10\,\mu$m, i.e. by 
four orders of magnitude less than that in the traditional acousto-optics. 
By using the dispersion equation (\ref{disp}) we derive the following 
estimate for $l_{\rm MC-ac}$ associated with the main 
acoustically-induced stop gap (one-phonon transition): 
\begin{equation}
l_{\rm MC-ac} \simeq { 4 \pi v^{\rm MC}_{\rm pol}(k/2) \over 
\Delta^{\rm MC}_{\rm ac}(I_{\rm ac},k/2) } \propto 
{ 1 \over I_{\rm ac} } \, , 
\eqnum{12}
\label{length}
\end{equation} 
where $\Delta^{\rm MC}_{\rm ac}$ is given by equation (\ref{gap}) and 
$v^{\rm MC}_{\rm pol}(p_{\|}) = \partial \omega^{\rm MC}_{\rm pol} /\partial 
p_{\|}$ is the MC polariton group velocity. For $I_{\rm SAW} = 4.5$\,mW/mm 
and $\nu_{\rm ac} = 2.28$\,GHz, equation (\ref{length}) yields 
$l_{\rm MC-ac} \simeq 13.3\,\mu$m for a zero-detuning GaAs-based microcavity 
with the Rabi energy given by $\hbar \Omega^{\rm MC}_{\rm X} = 3.7$\,meV 
(see figure\,1(b)). For the above values of $\nu_{\rm ac}$ and $\hbar 
\Omega^{\rm MC}_{\rm X}$ the relevant polariton group velocity is 
$v^{\rm MC}_{\rm pol}(k/2) \simeq 0.53 \times 10^8$\,cm/s. 

Two schemes ``SAW pumping - optical probing'' can be proposed for 
semiconductor microcavities: one deals with the use of bulk 
incoming/outgoing photons, which resonantly couple with MC polaritons 
through a MC Bragg reflector, and another involves in-plane light delivered 
to and collected from the lateral surfaces of a MC chip (see figure\,2). The 
evanescent acoustic field associated with a SAW decays nearly exponentially 
in the $z$-direction (the direction normal to the surface and, therefore, 
to the phonon wavevector ${\bf k}$) with a characteristic decay 
length $l^{z}_{\rm SAW}$ of a few $\lambda_{\rm ac}$. Note that for 
acoustically pumped GaAs-based microstructures the SAWs can usually be 
interpreted in terms of a Rayleigh wave \cite{Oliner78,Hess02}. For the 
SAW frequency $\nu_{\rm SAW} = 2.28$\,GHz, which has been used to calculate 
the plots shown in figures\,1(a) and 1(b), the phonon wavevector is $k = 5 
\times 10^4\,\mbox{cm}^{-1}$ so that $\lambda_{\rm ac} \simeq 1.3\,\mu$m. 
In this case $l^{z}_{\rm SAW} < l_{\rm MC-ac}$ and, therefore, the 
first scheme for the realization of the SAW-induced resonant acousto-optic 
effect is less favourable than the second one. 

The polariton effect in GaN-, ZnSe- and CdTe-based microcavities can be 
large, $\hbar \Omega^{\rm MC}_{\rm X} \gtrsim 20-30$\,meV 
\cite{Dang98,Boeuf00,Pawlis02,Malpuech02}. A self-consistent 
combination of the 
large binding energies of QW excitons and the large MC Rabi-splitting 
energies gives rise to the well-defined in-plane polariton eigenstates 
in the above microstructures even at the room temperature \cite{Pawlis02}. 
Thus the resonant acousto-optic Stark effect can also be realized in these 
structures at high temperatures. At the room temperature, however, the 
rate of incoherent scattering of QW excitons is larger than the radiative 
width of MC polaritons, i.e. $\gamma_{\rm X} > \Gamma_{\rm R}$. In this 
case the acoustic intensity threshold of the Stark effect is determined 
by $\hbar \Delta_{\rm ac}^{\rm MC}(I_{\rm ac}) \gtrsim \hbar \gamma_{\rm X}/2 
\sim 1$\,meV. 

The ``optical pumping - optical probing'' scheme mediated by acoustic 
phonons, i.e. the phonon-mediated optical Stark effect for QW excitons in 
microcavities, deals with the same three-particle exciton-phonon 
interaction as in the resonant acousto-optic effect. Therefore, a similar 
energy scale for the pump-induced spectral changes in both effects requires 
nearly the same concentrations of coherent virtual excitons 
$N_{\rm X}^{\rm MC} = [ \varphi^{\rm MC}({\tilde \omega}) I_{\rm opt}] / 
[v^{\rm MC}_{\rm pol}({\tilde \omega}) \hbar {\tilde \omega}]$ driven by 
the light field of intensity $I_{\rm opt}$ in the first case and coherent 
acoustic phonons $N^{\rm ph}_0 = I_{\rm ac}/(v_s \hbar 
\Omega^{\rm ac}_{\bf k})$ used in the latter one. Here ${\tilde \omega} 
\simeq \omega_0$ is the frequency of the pump optical wave, which induces a 
high-intensity MC polariton, and the excitonic component 
$\varphi^{\rm MC}({\tilde \omega})$ of the pump polariton is given by 
equation (\ref{comp}). Thus in order to generate $N_{\rm X}^{\rm MC} \sim 
N^{\rm ph}_0$ one needs much higher optical intensities than 
those necessary for the resonant acousto-optic Stark effect, i.e. 
$I_{\rm opt} \gg I_{\rm ac}$, because the MC polaritons have much higher 
frequency and lager group velocity than the acoustic phonons 
(${\tilde \omega} \gg \Omega_{\bf k}^{\rm ac}$ and $v^{\rm MC}_{\rm pol} 
> v_s$). For example, the intensity $I_{\rm SAW} = 4.5$\,mW/mm, which 
gives rise to the acoustically-induced spectrum plotted in figure\,1(b), 
corresponds to the average concentration of the coherent phonons, 
distributed in the surface layer of thickness $\lambda_{\rm ac}$, 
$\langle N_0^{\rm ph} \rangle \simeq 0.92 \times 10^{22}\,\mbox{cm}^{-3}$. 
A similar concentration of QW excitons definitely cannot be realized in GaAs 
nano-structures, due to complete removal of the excitonic states by the 
phase space filling effect.

\section{Acoustically-driven microcavities for optical 
modulation and switching}

For GaAs-based microcavities, the thickness of a top MC Bragg reflector 
is usually about $2-3\,\mu$m, i.e. is comparable with the SAW decay 
length $l^z_{\rm SAW}$ in the $z$-direction. In the meantime the in-plane 
interaction length $l_{\rm MC-ac}$ is compatible with the length scale 
needed for the effective work of an interdigital transducer (IT, a grid of 
equally-separated metalic stripes deposited on the surface of a 
piezoelectric crystal; an AC voltage applied to the IT induces a SAW). 
This makes the resonant acousto-optic effect in SAW-driven microcavities 
very attractive for possible applications \cite{Ivanov01b}. In figure\,2 
we show a possible design of a MC-based acousto-optic chip. The device can 
be used for optical modulation and switching \cite{Ivanov01b}. 

Because the main optical stop gap $\Delta^{\rm MC}_{\rm ac} \propto 
\sqrt{I_{\rm SAW}}$ follows the acoustic intensity, the dynamical response 
of the resonant acousto-optic Stark effect is mainly determined by a time 
required to build up the coherent acoustic field. The use of a SAW is 
favourable for the reduction of the above time, because the wave is driven 
through its electric component. In this case the piezoelectric coupling 
between excitons, the SAW, and an interdigital transducer reduces the 
switching on/off times of the SAW-induced Stark effect to a sub-nanosecond 
time scale.

\section{Conclusions}

In this paper we have developed resonant acousto-optics of MC polaritons. 
The following conclusions summarize our results. 

(i) The giant acousto-optic nonlinearities we propose and calculate are 
due to virtual QW excitons which resonantly mediate the interaction of 
coherent (surface) acoustic phonons with MC photons. 

(ii) The acoustically-induced Stark effect gives rise to the optical band 
gaps in the MC polariton dispersion. The gaps open up and develop with 
increasing intensity of the applied acoustic field and drastically change 
the reflection/transmission spectrum associated with MC polaritons. 

(iii) The resonant acousto-optic Stark effect can be realized in standard 
GaAs-based microcavities parametrically driven by a SAW of the frequency 
$\nu_{\rm ac} \lesssim 2-3\,$GHz and of the moderate intensity $I_{\rm SAW} 
\sim 0.1 - 10$\,mW/mm. The SAWs with the above characteristics are 
experimentally accessible. 

(iv) The use of the acoustic pump and optical probe practically does not 
lead to the generation of real excitations (QW excitons) and, therefore, 
to the increase of the scattering rates. In contrast, the latter effect, 
which is proportional to the density of photo-generated real, incoherent 
excitons and free carriers, considerably complicates a controlled 
realization of the optical Stark effects within the conventional scheme 
``optical pump - optical probe''. In the meantime the absence of real 
photo-excitations allows us to develop the rigorous, exactly-solvable 
model of the acousto-optic Stark effect in semiconductor microcavities. 

(v) The resonant acousto-optic effect in SAW-pumped microcavities 
is very promising for device applications such as tunable optical 
filters, optical modulators, etc.

\section{Acknowledgments}

We appreciate valuable discussions with L~V~Keldysh, L~E~Smallwood, 
A~V~Soroko, V~I~Talyanskii and E~A Zhukov. Support of this work by the 
EU RTN Project HPRN-CT-2002-00298 is gratefully acknowledged.

\begin{figure}
\centerline{\epsfxsize=8.4cm \epsfclipon
\epsffile{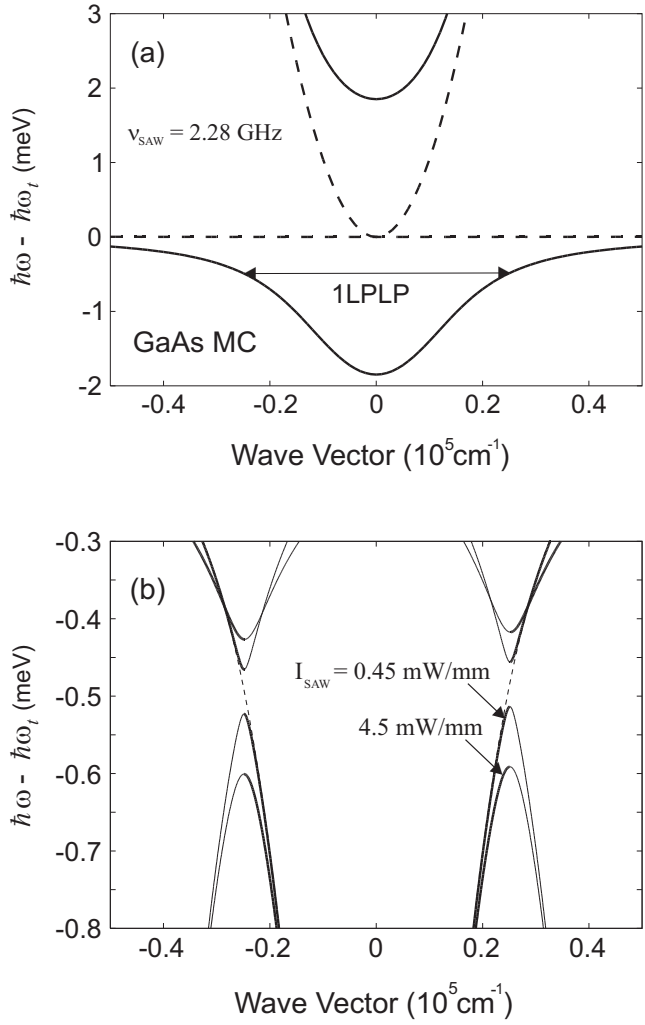}
}\vspace{3mm} 
\caption{ An acoustically-pumped GaAs-based microcavity with the Rabi 
splitting energy of MC polaritons $\hbar \Omega_{\rm X}^{\rm MC} = 
3.7\,$meV and zero detuning, i.e. $\omega_0 = \omega_t$. The (surface) 
acoustic wave is characterized by in-plane $|{\bf k}| = 5 \times 
10^4\,\mbox{cm}^{-1}$ and $\nu_{\rm ac} = \nu_{\rm SAW} = 2.28$\,GHz. 
(a) The in-plane QW 
exciton and MC photon dispersions (dashed lines), and the MC polariton 
dispersion branches (solid lines) calculated by equation (\ref{MC}). The 
arrow 1LPLP indicates one-phonon resonant transition between two 
counter-propagating lower-branch MC polaritons with in-plane wavevectors 
${\bf p}_{\|} = \pm {\bf k}/2$. (b) The acoustically-induced quasi-energy 
spectrum calculated by equation (\ref{disp}) for a spectral vicinity of 
the first optical band gap. The acoustic intensity is given by $I_{\rm SAW} 
= 0.45$ and 4.5\,mW/mm. The thick solid lines visualize the 
acoustically-induced gap in MC polariton spectrum. } 
\label{fig1}
\end{figure}

\begin{figure}
\centerline{\epsfxsize=8.4cm \epsfclipon
\epsffile{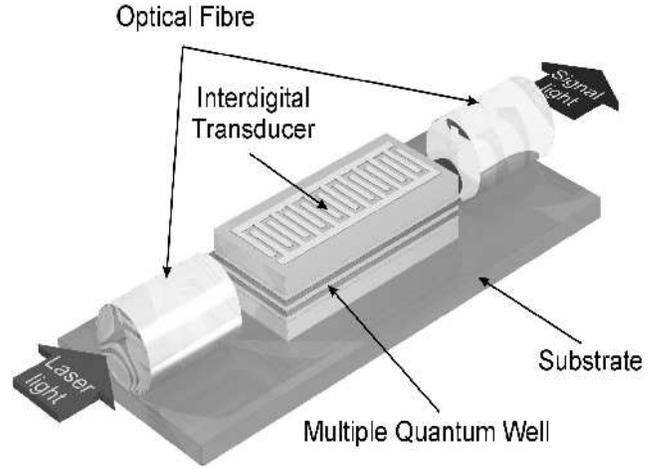}
}\vspace{3mm}
\caption{ Schematic of a SAW-driven semiconductor microcavity for optical 
modulation and switching. The surface acoustic wave with in-plane wavevector 
${\bf k}$ resonantly couples two counter-propagating MC polariton states, 
$\{ {\bf p}_{\|}$=${\bf k}/2,\,\omega^{\rm MC}(k/2) \}$ and 
$\{ {\bf p}_{\|}$=$-{\bf k}/2,\,\omega^{\rm MC}(k/2) \}$. The probe light 
field has frequency $\omega \simeq \omega^{\rm MC}(k/2)$. }
\label{fig2}
\end{figure}

\end{document}